\documentclass[aps,twocolumn,showpacs,preprintnumbers,amsmath,amssymb,superscriptaddress,prc]{revtex4-1}

\usepackage{graphicx}
\usepackage{dcolumn}
\usepackage{bm}
\usepackage{epsfig}
\usepackage{float}
\usepackage{amssymb}   

\makeatletter
\def\thickhline{%
  \noalign{\ifnum0=`}\fi\hrule \@height \thickarrayrulewidth \futurelet
   \reserved@a\@xthickhline}
\def\@xthickhline{\ifx\reserved@a\thickhline
               \vskip\doublerulesep
               \vskip-\thickarrayrulewidth
             \fi
      \ifnum0=`{\fi}}
\makeatother

\newlength{\thickarrayrulewidth}
\begin{document}


\title{$\beta$-Decay Half-Lives of 55 Neutron-Rich Isotopes beyond the $N$=82 Shell Gap}

\author{J.~Wu}%
\affiliation{%
Physics Division, Argonne National Laboratory, Argonne, Illinois 60439, USA}%
\affiliation{%
RIKEN Nishina Center, 2-1 Hirosawa, Wako-shi, Saitama 351-0198, Japan}%

\author{S.~Nishimura}%
\affiliation{%
RIKEN Nishina Center, 2-1 Hirosawa, Wako-shi, Saitama 351-0198, Japan}%

\author{P.~M\"oller}%
\affiliation{%
P. Moller Scientific Computing and Graphics, Inc., P. O. Box 1440, Los Alamos, NM 87544, USA}%

\author{M.~R.~Mumpower}%
\affiliation{%
Theoretical Division, Los Alamos National Laboratory, Los Alamos, New Mexico 87545, USA}%

\author{R.~Lozeva}%
\affiliation{%
IPHC, IN2P3/CNRS, University of Strasbourg, F-67037 Strasbourg Cedex 2, France}%
\affiliation{%
CSNSM, IN2P3/CNRS, University Paris-Saclay, F-91405 Orsay Campus, France}%

\author{C.~B.~Moon}%
\affiliation{%
Faculty of Science, Hoseo University, Chung-Nam 31499, Republic of Korea}%

\author{A.~Odahara}%
\affiliation{%
Department of Physics, Osaka University, Machikaneyama-machi 1-1, Osaka 560-0043 Toyonaka, Japan}%

\author{H.~Baba}%
\affiliation{%
RIKEN Nishina Center, 2-1 Hirosawa, Wako-shi, Saitama 351-0198, Japan}%

\author{F.~Browne}%
\affiliation{%
School of Computing, Engineering and Mathematics, University of Brighton, Brighton BN2 4GJ, United Kingdom}%
\affiliation{%
RIKEN Nishina Center, 2-1 Hirosawa, Wako-shi, Saitama 351-0198, Japan}%

\author{R.~Daido}%
\affiliation{%
Department of Physics, Osaka University, Machikaneyama-machi 1-1, Osaka 560-0043 Toyonaka, Japan}%

\author{P.~Doornenbal}%
\affiliation{%
RIKEN Nishina Center, 2-1 Hirosawa, Wako-shi, Saitama 351-0198, Japan}%

\author{Y.~F.~Fang}%
\affiliation{%
Department of Physics, Osaka University, Machikaneyama-machi 1-1, Osaka 560-0043 Toyonaka, Japan}%

\author{M.~Haroon}%
\affiliation{%
Department of Physics, Osaka University, Machikaneyama-machi 1-1, Osaka 560-0043 Toyonaka, Japan}%

\author{T.~Isobe}%
\affiliation{%
RIKEN Nishina Center, 2-1 Hirosawa, Wako-shi, Saitama 351-0198, Japan}%

\author{H.~S.~Jung}%
\affiliation{%
Department of Physics, Chung-Ang University, Seoul 06974, Republic of Korea}%

\author{G.~Lorusso}%
\affiliation{%
RIKEN Nishina Center, 2-1 Hirosawa, Wako-shi, Saitama 351-0198, Japan}%
\affiliation{%
National Physical Laboratory, NPL, Teddington, Middlesex TW11 0LW, United Kingdom}%
\affiliation{%
Department of Physics, University of Surrey, Guildford GU2 7XH, United Kingdom}%

\author{B.~Moon}%
\affiliation{%
Department of Physics, Korea University, Seoul 02841, Republic of Korea}%

\author{Z.~Patel}%
\affiliation{%
Department of Physics, University of Surrey, Guildford GU2 7XH, United Kingdom}%
\affiliation{%
RIKEN Nishina Center, 2-1 Hirosawa, Wako-shi, Saitama 351-0198, Japan}%

\author{S.~Rice}%
\affiliation{%
Department of Physics, University of Surrey, Guildford GU2 7XH, United Kingdom}%
\affiliation{%
RIKEN Nishina Center, 2-1 Hirosawa, Wako-shi, Saitama 351-0198, Japan}%

\author{H.~Sakurai}%
\affiliation{%
RIKEN Nishina Center, 2-1 Hirosawa, Wako-shi, Saitama 351-0198, Japan}%
\affiliation{%
Department of Physics, University of Tokyo, Hongo 7-3-1, Bunkyo-ku, 113-0033 Tokyo, Japan}%

\author{Y.~Shimizu}%
\affiliation{%
RIKEN Nishina Center, 2-1 Hirosawa, Wako-shi, Saitama 351-0198, Japan}%

\author{L.~Sinclair}%
\affiliation{%
Department of Physics, University of York, Heslington, York, YO10 5DD, U.K.}%
\affiliation{%
RIKEN Nishina Center, 2-1 Hirosawa, Wako-shi, Saitama 351-0198, Japan}%

\author{P.-A.~S\"oderstr\"om}%
\affiliation{%
RIKEN Nishina Center, 2-1 Hirosawa, Wako-shi, Saitama 351-0198, Japan}%

\author{T.~Sumikama}%
\affiliation{%
RIKEN Nishina Center, 2-1 Hirosawa, Wako-shi, Saitama 351-0198, Japan}%

\author{H.~Watanabe}%
\affiliation{%
IRCNPC, School of Physics and Nuclear Energy Engineering, Beihang University, Beijing 100191, China}%
\affiliation{%
RIKEN Nishina Center, 2-1 Hirosawa, Wako-shi, Saitama 351-0198, Japan}%

\author{Z.~Y.~Xu}%
\affiliation{%
Department of Physics, the University of Hong Kong, Pokfulam Road, Hong Kong}%
\affiliation{%
Department of Physics, University of Tokyo, Hongo 7-3-1, Bunkyo-ku, 113-0033 Tokyo, Japan}%

\author{A.~Yagi}%
\affiliation{%
Department of Physics, Osaka University, Machikaneyama-machi 1-1, Osaka 560-0043 Toyonaka, Japan}%

\author{R.~Yokoyama}%
\affiliation{%
Center for Nuclear Study (CNS), University of Tokyo, Wako-shi, Saitama 351-0198, Japan}%

\author{D.~S.~Ahn}%
\affiliation{%
RIKEN Nishina Center, 2-1 Hirosawa, Wako-shi, Saitama 351-0198, Japan}%

\author{F.~L.~Bello~Garrote}%
\affiliation{%
University of Oslo, P.O. Box 1072 Blindern, 0316 Oslo, Norway}%

\author{J.~M.~Daugas}%
\affiliation{%
CEA, DAM, DIF, F-91297 Arpajon Cedex, France}%

\author{F.~Didierjean}%
\affiliation{%
PHC, CNRS, IN2P3 and University of Strasbourg, F-67037 Strasbourg Cedex 2, France}%

\author{N.~Fukuda}%
\affiliation{%
RIKEN Nishina Center, 2-1 Hirosawa, Wako-shi, Saitama 351-0198, Japan}%

\author{N.~Inabe}%
\affiliation{%
RIKEN Nishina Center, 2-1 Hirosawa, Wako-shi, Saitama 351-0198, Japan}%

\author{T.~Ishigaki}%
\affiliation{%
Department of Physics, Osaka University, Machikaneyama-machi 1-1, Osaka 560-0043 Toyonaka, Japan}%

\author{D.~Kameda}%
\affiliation{%
RIKEN Nishina Center, 2-1 Hirosawa, Wako-shi, Saitama 351-0198, Japan}%

%

\author{I.~Kojouharov}%
\affiliation{%
GSI Helmholtzzentrum f\"ur Schwerionenforschung GmbH, 64291 Darmstadt, Germany}%

\author{T.~Komatsubara}%
\affiliation{%
RIKEN Nishina Center, 2-1 Hirosawa, Wako-shi, Saitama 351-0198, Japan}%

\author{T.~Kubo}%
\affiliation{%
RIKEN Nishina Center, 2-1 Hirosawa, Wako-shi, Saitama 351-0198, Japan}%

\author{N.~Kurz}%
\affiliation{%
GSI Helmholtzzentrum f\"ur Schwerionenforschung GmbH, 64291 Darmstadt, Germany}%

\author{K.~Y.~Kwon}%
\affiliation{%
Rare Isotope Science Project, Institute for Basic Science, Daejeon 34047, Republic of Korea}%

\author{S.~Morimoto}%
\affiliation{%
Department of Physics, Osaka University, Machikaneyama-machi 1-1, Osaka 560-0043 Toyonaka, Japan}%

\author{D.~Murai}%
\affiliation{%
RIKEN Nishina Center, 2-1 Hirosawa, Wako-shi, Saitama 351-0198, Japan}%

\author{H.~Nishibata}%
\affiliation{%
Department of Physics, Osaka University, Machikaneyama-machi 1-1, Osaka 560-0043 Toyonaka, Japan}%


\author{H.~Schaffner}%
\affiliation{%
GSI Helmholtzzentrum f\"ur Schwerionenforschung GmbH, 64291 Darmstadt, Germany}%


\author{T.~M.~Sprouse}%
\affiliation{%
Department of Physics, University of Notre Dame, Notre Dame, Indiana 46556, USA}%

\author{H.~Suzuki}%
\affiliation{%
RIKEN Nishina Center, 2-1 Hirosawa, Wako-shi, Saitama 351-0198, Japan}%

\author{H.~Takeda}%
\affiliation{%
RIKEN Nishina Center, 2-1 Hirosawa, Wako-shi, Saitama 351-0198, Japan}%

\author{M.~Tanaka}%
\affiliation{%
Research Center for Nuclear Physics (RCNP), Osaka University, Ibaraki, Osaka 567-0047, Japan}%

\author{K.~Tshoo}%
\affiliation{%
Rare Isotope Science Project, Institute for Basic Science, Daejeon 34047, Republic of Korea}%

\author{Y.~Wakabayashi}%
\affiliation{%
RIKEN Nishina Center, 2-1 Hirosawa, Wako-shi, Saitama 351-0198, Japan}%


\begin{abstract}
The $\beta$-decay half-lives of 55 neutron-rich nuclei $^{134-139}$Sn, $^{134-142}$Sb, $^{137-144}$Te, $^{140-146}$I, $^{142-148}$Xe, $^{145-151}$Cs, $^{148-153}$Ba, $^{151-155}$La were measured at the Radioactive Isotope Beam Factory (RIBF) employing the projectile fission fragments of $^{238}$U. The nuclear level structure, which relates to deformation, has a large effect on the half-lives. The impact of newly-measured half-lives on modeling the astrophysical origin of the heavy elements is studied in the context of $r$ process nucleosynthesis. 
For a wide variety of astrophysical conditions, including those in which fission recycling occurs, the half-lives have an important local impact on the second ($A$ $\approx$ 130) peak.
\end{abstract}

\date{\today}


\maketitle
The rapid neutron-capture ($r$-) process is responsible for creating about half of the stable nuclei beyond the isotopes of iron through a production mechanism of successive neutron captures intertwined with $\beta$ decays \cite{Burb57}. Recently there has been significant progresses in determining the astrophysical site(s) of $r$-process nucleosynthesis. Two promising $r$-process sites are a focus of discussion: the high-entropy neutrino wind of Type II supernovae \cite{Woosley94, Takahashi94, Qian96, Woosley92} and coalescence of neutron star binaries (NSNS) \cite{Lattimer74, Lattimer76, Symbalisty82, Eichler89, Davies94, Rosswog99}.  The historical first detection of Gravitational Waves (GW170817) from a binary neutron star merger has confirmed that the latter site produces $r$-process nuclei. The generation of an optical near-infrared transient source following a short Gamma-Ray Burst (GRB170817A), known as a kilonova, was believed to be produced by the radioactive decay of neutron-rich nuclei occurring in the nucleosynthesis of the $r$ process \cite{Abbott17, Drout17, Pian17, Hotokezaka18}. Although these recent observations constitute a valuable new contribution to $r$-process studies, more astronomical observations, experimental work and modeling will be required \cite{Frebel18}.

Obtaining the large amount of unmeasured nuclear properties that are necessary to calculate $r$-process nucleosynthesis is one of the main challenges in this endeavor. Fortunately, the critical input of $\beta$-decay half-lives are more readily accessed in experimental studies compared to other quantities, such as nuclear masses and neutron capture rates. Particularly challenging are nuclear structure issues occurring near closed shells, for example shape transitions from spherical to deformed nuclei and the appearance of octupole collectivity. Here, new measurements are of particular value. The newly measured $\beta$-delayed half-lives provide an important experimental data set to test the predictive ability of current theories.
 
In this paper, we present $\beta$-decay half-lives of 55 neutron-rich nuclei beyond the $N$ = 82 shell closure. Of these 55 nuclei, 13 are new with no data existing in the literature. The initial information on the nuclear structure in this region is investigated by comparing with different theoretical models. The impact of these new half-lives on the calculated $r$-process abundances near the ($A$ $\approx$ 130) peak is also explored.

\begin{center}
\begin{figure}[h!]
\resizebox{1.0\columnwidth}{!}{\rotatebox{0}{\includegraphics[clip=]{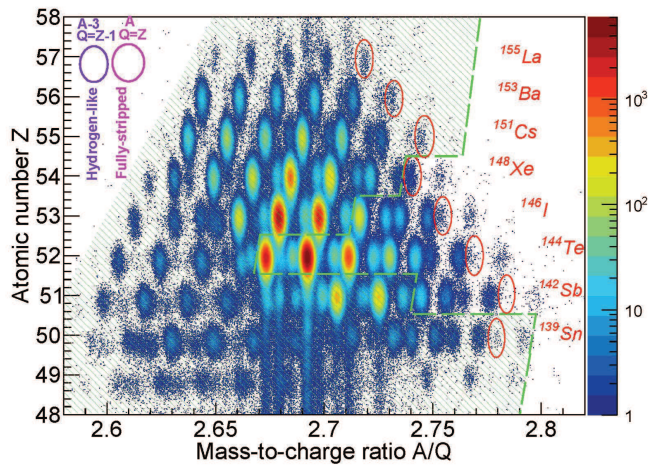}}}
\caption{\label{fig:chamber} (color online). Particle identification of the nuclides, and the well separated hydrogen-like charge state ($^{A-3}Z^{(Z-1)+}$) events accompanied closely on the left side of the fully-stripped ($^{A}Z^{Z+}$) isotopes as illustrated in the figure \cite{Shimizu18}. The isotopes included by the green shaded area were already measured in previous experiments. The nuclei tagged by red circles are the most exotic isotopes measured in this experiment for each element.} 
\end{figure}
\end{center}

The exotic neutron-rich isotopes around $^{140}$Te were produced at the Radioactive Isotope Beam Factory (RIBF), employing a primary beam of $^{238}$U with an energy of 345 MeV/u and an intensity of 5 pnA. After bombarding a target of $^{9}$Be with a thickness of 2.92 mm, the projectile fission fragments were selected and identified by the large-acceptance BigRIPS separator. The nuclei of interest were implanted in the beta counting system of the Wide range Active Silicon-Strip Stopper Array for Beta and ion detection (WAS3ABi) with a rate of about 100 pps \cite{Nishimura13}. The WAS3ABi includes a stack of five Double Sided Silicon Strip Detectors (DSSSDs) with 1 mm both for the width of each strip and the thickness of each DSSSD. The prompt and $\beta$-delayed $\gamma$ rays emitted from the implanted isotopes were detected by the 84 germanium cluster detectors of the Euroball RIken Cluster Array (EURICA), which surrounded the WAS3ABi \cite{Nishimura12, PA13, Wu15}. The particle identification plot was constructed by the TOF-B$\rho$-$\Delta$E principle with the measurements from the detectors along the beam line of the BigRIPS separator, including plastic scintillators, ionization chambers, as well as a variety of deflection magnets \cite{Fukuda13}. contaminant events with hydrogen-like charge states are well separated from the fully-stripped isotope events, which allow us to obtain the precise results employing both the charge-state and fully-stripped events \cite {Shimizu18} (see Fig.~\ref{fig:chamber}).

\begin{center}
\begin{figure}[h!]
\resizebox{1.0\columnwidth}{!}{\rotatebox{0}{\includegraphics[clip=]{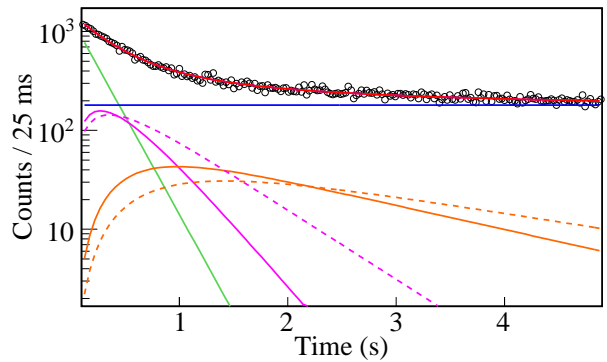}}}
\caption{(color online). Time distribution of $^{142}$Te $\beta$-decay events fitted to the sum of activities of several components: parent nuclei (solid green line), daughter nuclei (solid magenta line), granddaughter nuclei (solid orange line), $\beta$-delayed daughter nuclei (dashed magenta line), $\beta$-delayed granddaughter nuclei (dashed orange line), as well as a constant background (solid blue line). }  
\end{figure}
\end{center}

\begin{center}
\begin{figure}[h!]
\resizebox{1.0\columnwidth}{!}{\rotatebox{0}{\includegraphics[clip=]{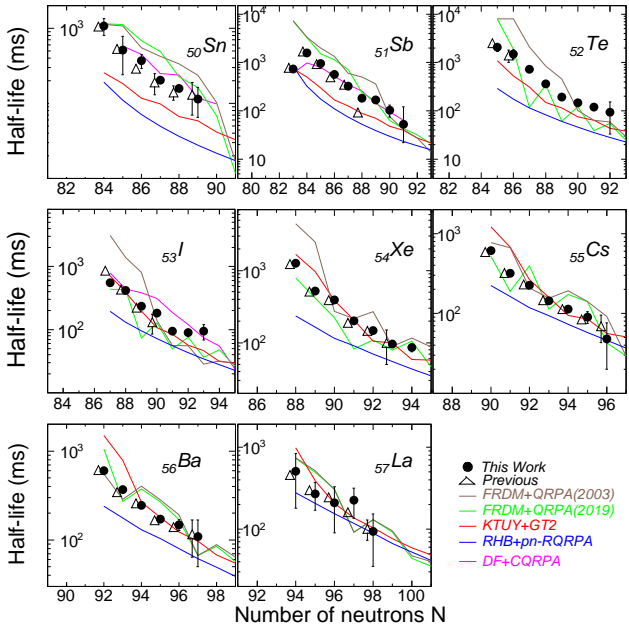}}}
\caption{\label{fig:sys} (color online). Systematic trends of the $\beta$-decay half-lives from this work (solid circles) and from previous measurements (open triangles) \cite{nndc, Lorusso15, Rudstam93, Arndt11, Samri85, Asghar75, Ahrens76, Aleklett80, Kessler06, Bergmann03, Wu17}. The measurements are compared to predictions of five theoretical models: FRDM+QRPA(2003) \cite{Moller03} (brown), FRDM+QRPA(2019) \cite{Moller18} (green), KTUY+GT2 \cite{Koura05, Tachibana90} (red), RHB+$pn$-RQRPA \cite{Marketin16} (blue), and DF+CQRPA \cite{Borzov16, Borzov08} (magenta).} 
\end{figure}
\end{center}

\begin{table*}[ht!]
\flushleft

\caption{\label{tab:data} $\beta$-decay half-lives measured in this work compared with literature values, when available. The half-lives of $^{140}$Sb and $^{140}$Te were recently published based on the dataset from the same experiment, with the values of 173 $\pm$ 12 \cite{Moon1795} and 350 $\pm$ 5 \cite{Moon1796}, respectively. The reason for small discrepancies probably comes from the background which the previous measurements didn't subtract while gating on the $\beta$-delayed $\gamma$ rays. The newly measured half-life of $^{139}$Sb has a large discrepancy compared with the previous measurement, probably because the old measurement suffers from the large background level in the decay curve \cite{Arndt11}. 
The nuclei with the half-lives confirmed by the method of gating the $\beta$-delayed $\gamma$ rays are tagged with an asterisk (*).} 
\setlength{\tabcolsep}{20pt}
\renewcommand{\arraystretch}{1.2}
\begin{tabular*}{1.0\textwidth}{ c   c   c   c   c   c}
\thickhline
\hline\hline
Nucleus & T$^{exp}_{1/2}$ (ms) & T$^{lit}_{1/2}$ (ms) & Nucleus & T$^{exp}_{1/2}$ (ms) & T$^{lit}_{1/2}$ (ms) \\
\thickhline
$^{134}$Sn                &  1070 $\pm$ 270          &       1050 $\pm$ 11 \cite{nndc}            &         $^{145}$I$^{\ast}$                   &  89.7 $\pm$ 9.3    &       ...                                      \\
$^{135}$Sn                &  510 $\pm$ 270           &       530 $\pm$ 20 \cite{nndc}              &         $^{146}$I                            &  94 $\pm$ 26       &       ...                                      \\
$^{136}$Sn                &  369 $\pm$ 76            &       290 $\pm$ 20 \cite{nndc}             &         $^{142}$Xe                           &  1260 $\pm$ 70     &       1230 $\pm$ 20 \cite{nndc}                \\
$^{137}$Sn                &  204 $\pm$ 12            &       230 $\pm$ 30 \cite{Lorusso15}        &         $^{143}$Xe$^{\ast}$                  &  519 $\pm$ 86      &       511 $\pm$ 6 \cite{Bergmann03}            \\
$^{138}$Sn                &  158 $\pm$ 15            &       140$^{+30}_{-20}$ \cite{Lorusso15}   &         $^{144}$Xe$^{\ast}$                  &  391 $\pm$ 52      &       388 $\pm$ 7 \cite{Ahrens76}              \\
$^{139}$Sn                &  114 $\pm$ 49            &       130 $\pm$ 60 \cite{Lorusso15}        &         $^{145}$Xe$^{\ast}$                  &  202 $\pm$ 26      &       188 $\pm$ 4 \cite{Bergmann03}            \\
$^{134}$Sb$^{\ast}$       &  730 $\pm$ 110           &       780 $\pm$ 60 \cite{nndc}             &         $^{146}$Xe$^{\ast}$                  &  147 $\pm$ 13      &       146 $\pm$ 6 \cite{Bergmann03}            \\
$^{135}$Sb$^{\ast}$       &  1570 $\pm$ 230          &       1679 $\pm$ 15 \cite{nndc}            &         $^{147}$Xe$^{\ast}$                  &  88 $\pm$ 14       &       100$^{+100}_{-50}$ \cite{Bergmann03}     \\
$^{136}$Sb$^{\ast}$       &  957 $\pm$ 79            &       923 $\pm$ 14 \cite{Rudstam93}        &         $^{148}$Xe                           &  85 $\pm$ 15       &       ...                                      \\
$^{137}$Sb$^{\ast}$       &  566 $\pm$ 52            &       492 $\pm$ 25 \cite{Arndt11}          &         $^{145}$Cs$^{\ast}$                  &  612 $\pm$ 20      &       587 $\pm$ 5 \cite{nndc}                  \\
$^{138}$Sb$^{\ast}$       &  326 $\pm$ 8             &       348 $\pm$ 15 \cite{nndc}             &         $^{146}$Cs$^{\ast}$                  &  318 $\pm$ 18      &       322.0 $\pm$ 1.3 \cite{nndc}              \\
$^{139}$Sb$^{\ast}$       &  182 $\pm$ 9             &       93$^{+14}_{-3}$ \cite{Arndt11}       &         $^{147}$Cs$^{\ast}$                  &  225 $\pm$ 5       &       230 $\pm$ 1 \cite{nndc}                  \\
$^{140}$Sb$^{\ast}$       &  169 $\pm$ 7             &       ...                                  &         $^{148}$Cs$^{\ast}$                  &  144 $\pm$ 9       &       146 $\pm$ 6 \cite{nndc}                  \\
$^{141}$Sb                &  103 $\pm$ 29            &       ...                                  &         $^{149}$Cs$^{\ast}$                  &  113 $\pm$ 6       &       113 $\pm$ 8 \cite{Wu17}                  \\
$^{142}$Sb                &  53$^{+69}_{-31}$        &       ...                                  &         $^{150}$Cs                           &  90 $\pm$ 15       &       84.4 $\pm$ 8.2 \cite{Wu17}               \\
$^{137}$Te                &  2080 $\pm$ 400          &       2490 $\pm$ 50 \cite{Samri85}         &         $^{151}$Cs                           &  48 $\pm$ 28       &       69 $\pm$ 26 \cite{Wu17}                  \\
$^{138}$Te                &  1500 $\pm$ 320          &       1400 $\pm$ 400 \cite{Asghar75}       &         $^{148}$Ba                           &  602 $\pm$ 46      &       612 $\pm$ 17 \cite{nndc}                 \\
$^{139}$Te$^{\ast}$       &  724 $\pm$ 81            &       ...                                  &         $^{149}$Ba                           &  368 $\pm$ 19      &       344 $\pm$ 7 \cite{nndc}                  \\
$^{140}$Te$^{\ast}$       &  360 $\pm$ 21            &       ...                                  &         $^{150}$Ba                           &  245 $\pm$ 16      &       259 $\pm$ 5 \cite{Wu17}                  \\
$^{141}$Te$^{\ast}$       &  193 $\pm$ 16            &       ...                                  &         $^{151}$Ba                           &  166 $\pm$ 11      &       167 $\pm$ 5 \cite{Wu17}                  \\
$^{142}$Te$^{\ast}$       &  147 $\pm$ 8             &       ...                                  &         $^{152}$Ba                           &  148 $\pm$ 21      &       139 $\pm$ 8 \cite{Wu17}                  \\
$^{143}$Te                &  120 $\pm$ 8             &       ...                                  &         $^{153}$Ba                           &  109 $\pm$ 59      &       116 $\pm$ 52 \cite{Wu17}                 \\
$^{144}$Te                &  93 $\pm$ 60             &       ...                                  &         $^{151}$La                           &  510 $\pm$ 330     &       457$^{+30}_{-18}$ \cite{Wu17}            \\
$^{140}$I$^{\ast}$        &  553 $\pm$ 46            &       860 $\pm$ 40 \cite{Ahrens76}         &         $^{152}$La                           &  270 $\pm$ 100     &       298$^{+6}_{-23}$ \cite{Wu17}             \\
$^{141}$I$^{\ast}$        &  418 $\pm$ 8             &       430 $\pm$ 20 \cite{Aleklett80}       &         $^{153}$La                           &  210 $\pm$ 120     &       245 $\pm$ 18 \cite{Wu17}                 \\
$^{142}$I$^{\ast}$        &  235 $\pm$ 11            &       222 $\pm$ 12 \cite{Kessler06}        &         $^{154}$La                           &  221 $\pm$ 89      &       161 $\pm$ 15 \cite{Wu17}                 \\
$^{143}$I$^{\ast}$        &  182 $\pm$ 8             &       130 $\pm$ 45 \cite{Kessler06}        &         $^{155}$La                           &  94 $\pm$ 59       &       101 $\pm$ 28 \cite{Wu17}                 \\
$^{144}$I$^{\ast}$        &  94 $\pm$ 8              &       ...                                  &                                                                                                                    \\

\hline\hline
\thickhline   
\end{tabular*}

\flushleft
  \label{tabN82}
  \end{table*}

\begin{center}
\begin{figure}
\resizebox{1.0\columnwidth}{!}{\rotatebox{0}{\includegraphics[clip=]{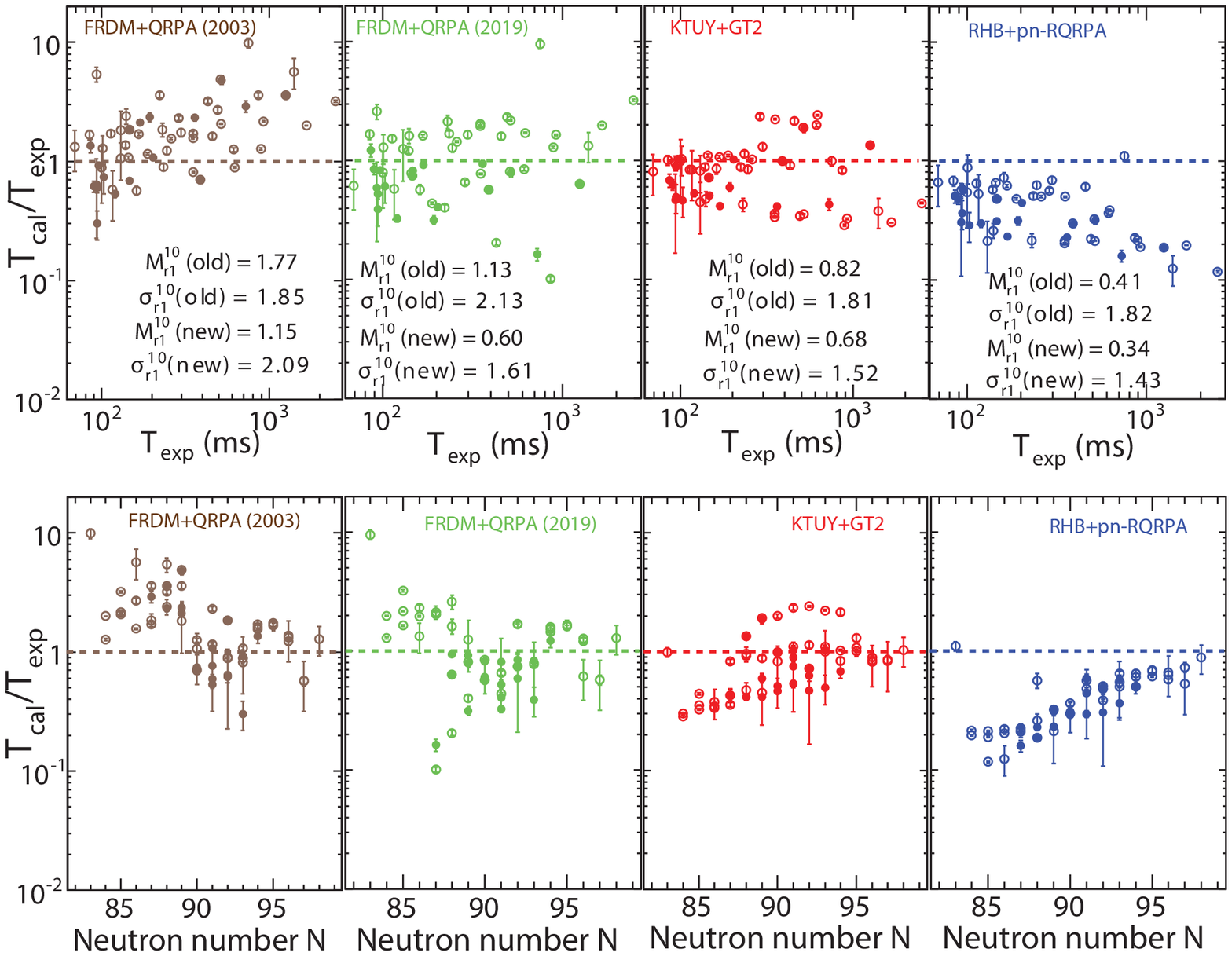}}}
\caption{(color online). Ratio between theoretical calculations and measured half-lives plotted versus experimental half-life and neutron number for the four models: FRDM+QRPA(2003) \cite{Moller03} (brown), FRDM+QRPA(2019) \cite{Moller18} (green), KTUY+GT2 \cite{Koura05, Tachibana90} (red), and RHB+$pn$-RQRPA \cite{Marketin16} (blue), with the previous measurements (empty dots) and 13 newly-measured half-lives (solid dots).}
\end{figure}
\end{center}

The $\beta$-decay half-life of each isotope was extracted by fitting its $\beta$-decay curve considering the time difference between the position-correlated implanted ions and its $\beta$ decays \cite{Xuphd, Nishimura11, Lorusso15, Xu14}. The fitting employed two major techniques: the least-squared and unbinned maximum likelihood methods. Both include the contributions from the decays of daughter and granddaughter nuclei, $\beta$-delayed daughter and granddaughter nuclei, as well as a constant background. An example of the decay curve fitting for $^{142}$Te is given in Fig.~2. The half-lives of descendant nuclei used in the fitting were measured in this experiment or taken from the literature \cite{nndc}. The $\beta$-delayed neutron emission probabilities ($P_{n}$) were obtained from literature values \cite{nndc} if they were known, otherwise they were regarded as free parameters with a range between zero and one. The main source of uncertainty comes from the statistical error, while the uncertainty from descendant nuclei and $P_{n}$ values were included as the systematic error. If enough statistics were collected, the obtained half-life was confirmed by fitting a decay curve which was gated on the $\beta$-delayed $\gamma$ rays. In this region, it is possible for the $\beta$ decay to be populated by the long-lived isomeric states. The isomers were identified by analyzing the $\beta$-decay schemes in detail when there were enough statistics. The half-lives of ground-state decays were extracted by gating on the $\beta$-delayed gamma-rays only populated from the ground states, such as $^{134}$Sb, $^{140}$I in this region. 
In this case, a large isomeric ratio and high statistics were necessary to disentangle the $\beta$-decays from the ground and isomeric states.
The detailed information will be published in a forthcoming publication. 

The measured half-lives are reported in Table I together with previous measurements, if available. The systematic trends of $\beta$-decay half-lives for exotic isotopes of elements from $_{50}$Sn to $_{57}$La are shown in Fig.~\ref{fig:sys}. The experimental results are compared with literature values \cite{nndc, Lorusso15, Rudstam93, Arndt11, Samri85, Asghar75, Ahrens76, Aleklett80, Kessler06, Bergmann03, Wu17} and five theoretical predictions: finite-range droplet-model FRDM-1995 mass formula \cite{Moller95} with quasi-particle-random-phase approximation (QRPA) (2003) \cite{Moller03} as well as the new version (2019) \cite{Moller18} with FRDM-2012 masses \cite{Moller16}, Koura-Tachibana-Uno-Yamada (KTUY) with the second generation of $\beta$-decay gross theory (GT2) \cite{Koura05, Tachibana90}, Relativistic Hartree-Bogoliubov (RHB) with the proton-neutron Relativistic quasiparticle random phase approximation ($pn$-RQRPA) \cite{Marketin16}, and the energy density functional (DF) with continuum quasi-particle random-phase approximation (CQRPA) \cite{Borzov16, Borzov08}. 

The comparison (Fig.~3) indicates that the current theoretical models are unable to fully reproduce the experimental results in general up to a factor of ten in this region. The global nature of the KTUY+GT2 calculation typically does better for the neutron-rich isotopes of elements beyond the $Z$ = 50 shell gap (see $_{53}$I, $_{54}$Xe, $_{55}$Cs, $_{56}$Ba, $_{57}$La). However, it cannot reproduce the data for the nearly spherical nuclei close to the shell gap since non-GT transitions contribute, for example a first forbidden decay $\nu f_{7/2}$ $\Rightarrow$ $\pi g_{9/2}$ in $N$ $>$ 82 nuclei near the doubly-magic nuclei $^{132}$Sn \cite{Lorusso15}. The systematic trends of $\beta$-decay half-lives in the rare-earth region have been well understood based on the FRDM+QRPA calculation \cite{Wu17}. However, in this region, it shows relatively large odd-even staggering caused by the nucleon pairing effect, and tends to show an overestimation when approaching $_{50}$Sn. The use of FRDM-2012 masses does improve the agreement in this region especially for $_{50}$Sn, $_{51}$Sb and $_{52}$Te (see green curve Fig.~\ref{fig:sys}). The RHB+$pn$-QRPA does not reproduce any structure-related features including the odd-even staggering, and generally underestimates the experimental half-lives. One reason for the discrepancy is that the $Q_{\beta}$ values in this calculation do not include pairing effects. The DF+CQRPA are only applicable for nearly spherical nuclei, so they fit well the half-life trends of $_{50}$Sn and $_{51}$Sb, which are close to the doubly-magic nucleus $^{132}_{50}$Sn.

The KTUY+GT2 and RHB+$pn$-RQRPA, systematically underestimate the half-lives but agree better for more neutron-rich nuclei further from stability, because for large $Q_{\beta}$ half-lives are less affected by level-structure inaccuracies. Although the two FRDM+QRPA calculations are based on detailed calculated single-particle spectra, they are less accurate close to $N$ = 82, which may be related to the difficulty of describing transition regions between spherical and well-deformed nuclei. All models become more accurate further from stability, because of the large $Q_{\beta}$ the half-lives are less sensitive to small inaccuracies in level structure (see Ref.~\cite{Moller18} for an in-depth discussion). A more quantitative comparison was carried out by calculating the ratio between theoretical predictions and experimental results as functions of experimental half-lives and neutron number $N$ (see Fig. 4). 
In terms of quantitative analysis \cite{Moller16}, the mean deviation $M_{r1}^{10}=10^{M_{r1}}$ and mean fluctuation $\sigma_{r1}^{10}=10^{\sigma_{r1}}$ have been evaluated for the previous measurements and newly-measured data in this work, with $r=T_{exp}/T_{cal}$, $r1=log_{10}(r)$, $M_{r1}=1/n\sum_{i=1}^{n}r_{1}^{i}$, $\sigma_{r1}=(1/n\sum_{i=1}^{n}(r_{1}^{i}-M_{r1})^{2})^{1/2}$.
The deviations can be up to one-order of magnitude. The Mean Deviation $M_{r1}^{10}$ and Mean Fluctuation $\sigma_{r1}^{10}$ of the four theoretical predictions have been evaluated including both previous measurements and the newly-measured half-lives. For newly-measured half-lives, the FRDM+QRPA (2003) has a $M_{r1}^{10}$ value of 1.15, and a $\sigma_{r1}^{10}$ value of 2.09 which is similar to what was obtained for known nuclei \cite{Moller18}. The FRDM+QRPA (2019) and KTUY+GT2  have similar vales of $M_{r1}^{10}$ and $\sigma_{r1}^{10}$ indicating similar predictive accuracy in this region. The RHB+$pn$-QRPA substantially and systematically  underestimates the experimental values with a $M_{r1}^{10}$ value of 0.34, as is also evident from the figures.

\begin{center}
\begin{figure}
\resizebox{1.2\columnwidth}{!}{\rotatebox{0}{\includegraphics[clip=]{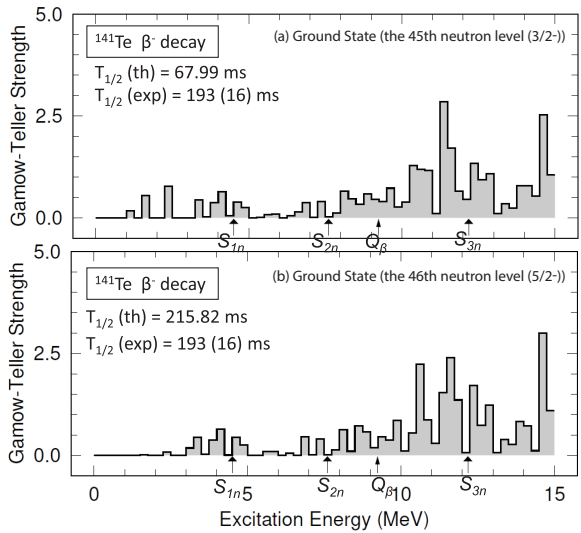}}}
\caption{Gamow-Teller strength functions in the daughter states following $\beta$-decay of $^{141}$Te calculated by the FRDM+QRPA(2019) model, based on the odd particle located in level 45 (a) and level 46 (b). The thin vertical arrow indicates the Q$_{\beta}$ value, the slightly thicker arrow indicates the neutron separation energy.} 
\end{figure}
\end{center}

\begin{center}
\begin{figure}
\resizebox{1.0\columnwidth}{!}{\rotatebox{0}{\includegraphics[clip=]{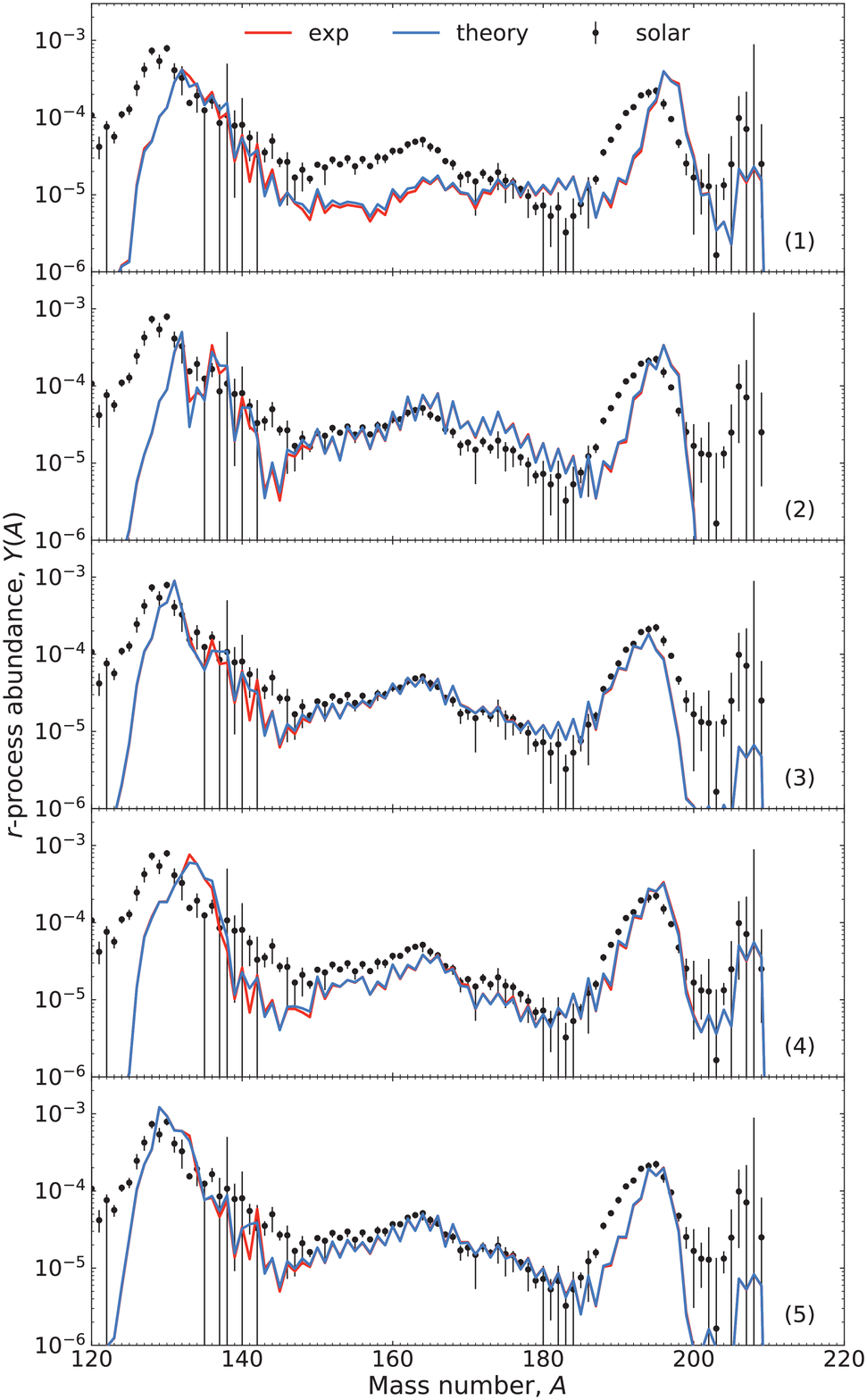}}}
  \caption{\label{fig:rprocess} (color online) The calculated $r$-process abundance with (red) and without (blue) the 13 newly-measured half-lives, by employing the five distinct astrophysical trajectories ((1) Low Entropy ($s$ $\approx$ 50), Initial Electron Fraction ($Y_{e}$=0.27). \cite{Holmbeck18} (2) High Entropy ($s$ $\approx$ 200), Initial Electron Fraction ($Y_{e}$ = 0.3). \cite{Mumpower16} (3) Low Entropy ($s$ $\approx$ 30), Initial Electron Fraction (artificially reduced to produce a main $r$-process, $A>$120). \cite{Arcones07} (4) Low Entropy ($s$ $\approx$ 10), Initial Electron Fraction ($Y_{e}$ = 0.05), with fission recycling. \cite{Goriely11} (5) Low Entropy ($s$ $\approx$ 10), Initial Electron Fraction ($Y_{e}$=0.19). \cite{Holmbeck18}) The experimental data were input if they are known, otherwise the theoretical properties were employed (Nuclear Mass (FRDM-2012 \cite{Moller16}) + T$_{1/2}$ and P$_{n}$ (FRDM+QRPA (2019) \cite{Moller18}) + ($n$,$\gamma$) rates (calculated based on FRDM-2012 \cite{Moller16}))The black points represent the solar $r$-process abundance. \cite{Goriely99}}
\end{figure}
\end{center}


Although the FRDM+QRPA (2019) has more consistent predictions than other models it exhibits large odd-even staggering in contrast to the experiment. Pairing effects are expected to give rise to substantial odd-even staggering, but specifics of the nuclear level structure can modify this expected general behavior \cite{Moller18}. The calculated half-lives of the $_{52}$Te isotopes with even neutron number agree well with experiment, but the odd-$N$ nuclei are underestimated. One possibility is that the calculated order of levels is not completely correct. We have investigated the effect of a small change in level order. 
Fig.~5 shows the calculated half-lives of $^{141}$Te$_{89}$ with two different odd-neutron configurations. When the position of the odd neutron, expected to be level 45 with spin-parity 3/2$^{-}$, is changed to the 46th neutron level with spin and parity of 5/2$^{-}$, the calculated half-life is much larger and more consistent with the experimental value. This means that in the actual $^{141}$Te the positions of these levels may be inverted with respect to the calculated ones. The change in spin/parity of the odd particle obviously leads to different transition strengths and energies resulting in the longer calculated half-life we obtain, see Fig.~5. A more complete discussion of this type of issues are in Ref.~\cite{Moller18}.

The impact of the measured half-lives on $r$-process abundances is explored using the PRISM (Portable Routines for Integrated nucleoSynthesis Modeling) reaction network \cite{Mumpower17} which supports unique control over nuclear physics inputs allowing for the variation of mass models, half-lives and fission properties \cite{Cote18}. Network calculations were performed using a range of five distinct conditions that could occur in an astrophysical environment to produce a full abundance pattern out to the third peak. Nuclear properties including nuclear masses, $\beta$-decay half-lives, neutron-capture rates in our network calculations, are predicted based on FRDM-2012 \cite{Moller16}. Experimental and evaluated data are used when available. In general, $\beta$ decay rates, $\lambda_{\beta}$, influence $r$-process abundances $Y$ along the $r$-process path through the steady beta-flow condition ($\lambda_{\beta}(Z,A_{path})Y(Z,A_{path})\approx$ constant) in the early stage, and during the decay back to stability in the late stages, while $\beta$ decay competes with neutron capture, see e.g. Ref.~\cite{Mumpower14} and references therein. 
The presently measured nuclei operate during the second phase, as the $r$-process path proceeds back to stability.
Figure~6 shows the difference of calculated $r$-process abundances with and without inclusion of the thirteen newly-measured half-lives of this work. This figure demonstrates the important regional influence of our newly measured half-lives compared to the baseline theoretical predictions. A quantitative discussion of improvement to the solar isotopic pattern is not clear due to the variation seen in this region among differing astrophysical conditions. Instead, we report the average impact factor, $F_{\rm avg} \sim 4.86$, and its standard deviation, $\sigma(F) \approx 1.22$, for the five astrophysical trajectories. The impact factor of this value indicates a strong local dependence on the final abundances \cite{Mumpower16}. For some of the trajectories, (1), (2), and (4), a tiny global effect up to the $A \approx 195$ $r$-process peak is observed owing to material that is cleared out from the second $r$-process peak \cite{Moller16,Moller18}. We also note that conditions with fission recycling do not qualitatively change this conclusion, as indicated by trajectory (5) in Fig.~6.

In summary, our decay measurement campaign has reached the furthest from the stability line, now beyond $N$ = 82 shell gap. The newly measured half-lives were found to have an important local impact on the abundances of the second ($A$ $\approx $130) $r$-process peak. The comparison with measured half-lives illustrates the performance of various theoretical predictions, and indicates that the order of orbitals, which is sensitive to the nuclear deformation in the transition region, have a large impact on the calculated half-lives. The $r$-process calculations here do not fully reproduce  all the peaks in the $r$-process abundance pattern. 
A more comprehensive understanding of the $r$-process and its astrophysical sites are still the goal we are continuing to pursue. Future studies which probe more exotic neutron-rich nuclei will aid in this endeavor.

%
This work was carried out at the RIBF operated by RIKEN Nishina Center, RIKEN, CNS, University of Tokyo. The authors acknowledge discussion with Dr. Calem Hoffman (ANL). We also acknowledge the EUROBALL Owners Committee for the loan of germanium detectors and the PreSpec Collaboration for the readout electronics of the cluster detectors. Part of the WAS3ABi was supported by the Rare Isotope Science Project which is funded by the Ministry of Education, Science, and Technology (MEST) and National Research Foundation (NRF) of Korea (2013M7A1A1075764). This work was supported by the U.S. Department of Energy, Office of Science, Office of Nuclear Physics, at Argonne National Laboratory under Contract No. DE-AC02-06CH11357, KAKENHI (Grants No. 25247045, 17H06090), the RIKEN Foreign Research Program, the Spanish Ministerio de Ciencia e Innovacin (Contracts No. FPA2009-13377-C02 and No. FPA2011-29854-C04), the UK Science and Technology Facilities Council, the National Research Foundation Grant funded by the Korean Government (Grants No. NRF-2009-0093817, No. NRF-2015R1D1A1A01056918, No. NRF-2016R1A5A1013277 and No. NRF-2013R1A1A2063017), and the Hungarian Scientific Research Fund OTKA Contract No. K100835. M.R. Mumpower carried out this work under the auspices of the National Nuclear Security Administration of the U.S. Department of Energy at Los Alamos National Laboratory under Contract No. DE-AC52-06NA25396. 
%

\bibliography{refs}

\end{document}